\documentclass{article}

\usepackage{arxiv}

\usepackage[utf8]{inputenc} 
\usepackage[T1]{fontenc}    
\usepackage{hyperref}       
\usepackage{url}            
\usepackage{booktabs}       
\usepackage{amsfonts}       
\usepackage{nicefrac}       
\usepackage{microtype}      
\usepackage{lipsum}		
\usepackage{graphicx}
\usepackage{natbib}
\usepackage{doi}
\usepackage{appendix}
\usepackage{graphicx}
\usepackage{algorithm}
\usepackage{algpseudocode}

\title{An Empirical Comparison of the Quadratic Sieve Factoring Algorithm and the Pollard Rho Factoring Algorithm}

\author{ Zongxia Li
     \\
	Computer Science and Mathematics\\
	University of Maryland\\
	College Park, MD 20742 \\
	\texttt{zli12321@umd.edu} \\
	\And
	\href{https://www.cs.umd.edu/~gasarch/}{\includegraphics[scale=0.06]{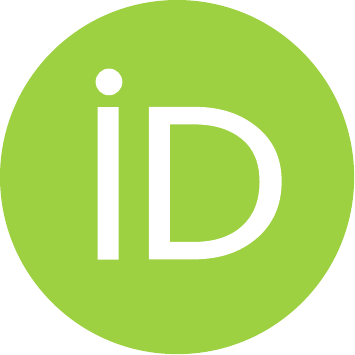}\hspace{1mm}William Gasarch} \thanks{Use footnote for providing further
		information about author (webpage, alternative
		address)---\emph{not} for acknowledging funding agencies.}\\
	Computer Science Affiliate in Mathematics\\
	University of Maryland\\
	College Park, MD 20742 \\
	\texttt{gasarch@umd.ed} \\
}

\renewcommand{\shorttitle}{Technical Report}

\hypersetup{
pdftitle={A template for the arxiv style},
pdfsubject={q-bio.NC, q-bio.QM},
pdfauthor={David S.~Hippocampus, Elias D.~Striatum},
pdfkeywords={First keyword, Second keyword, More},
}

\begin{document}
\maketitle

\begin{abstract}
One of the most significant challenges on cryptography today is the problem of factoring large integers since there are no algorithms that can factor in polynomial time, and factoring large numbers more than some limits(200 digits) remain difficult. The security of the current cryptosystems depends on the hardness of factoring large public keys. In this work, we want to implement two existing factoring algorithms - pollard-rho and quadratic sieve - and compare their performance. In addition, we want to analyze how close is the theoretical time complexity of both algorithms compared to their actual time complexity and how bit length of numbers can affect quadratic sieve's performance. Finally, we verify whether the quadratic sieve would do better than pollard-rho for factoring numbers smaller than 80 bits. 
\end{abstract}


\section{Introduction}
The idea of public-key cryptography was first introduced in 1975 by Martin Hellman, Ralph Merkle, and Whitfield Diffie at Stanford University~\cite{ten_years}. Before the public key cryptosystem era, if two people want to exchange secret information without anybody else knowing, they have to agree in advance on a secret key known only by them but not anyone else. After the invention of public-key systems, two people could exchange secret information without ever meeting each other. The idea is that the secret message can only be decrypted in a reasonable amount of time using secret keys possessed by two people exchanging information. Then Ron Rivest, Adi Shamir, and Leonard Adleman introduced the RSA public-key cryptosystem that is considered more secure than previous cryptosystems. This cryptosystem is implemented based on two ideas: public-key encryption and digital signatures~\cite{RSA}. In the public-key encryption part, the sender generates a random prime p, a base number g, and a random number a.

Then, the sender sends need to send (p, g, g$^a$( (mod p)) to the receiver. After the receiver receives the number, the receiver generates a random number b, and send g$^b$ (mod p) back to the sender. The sender and the receiver can then compute the shared secret key easily based on the information they have ~\cite{diffie}. However, for a third party to know the secret, they have to compute g$^{ab}$ with four values - p, g, g$^a$( (mod p), and g$^b$ (mod p). The Diffie-Hellman key exchange algorithm is the place where the factoring algorithm comes in. If the public key consists of two large primes, where each prime is roughly 512 bits, or 1024 bits, or 2048 bits, then the third party would have to find the factors for the public key to find out the secret key. There are already many existing factoring algorithms. However, none of them can factor large products in a reasonable time- polynomial time. Nowadays, it is feasible to factor 155- decimal digit numbers, but it is still considered hard to factor numbers more than 150 decimal digits ~\cite{framework}. Although we cannot examine the runtime of those algorithms on large products- more than 512 bits, we can analyze some of those algorithms with relatively small scale numbers- numbers fewer than 100 bits. Pollard-rho and quadratic-sieve are two popular factoring algorithms we analyze in this work.

We implement the pollard-rho algorithm based on the birthday paradox, and probability theory ~\cite{Rho}. Then we implement the quadratic sieve algorithm and optimize some important steps using fast Gaussian elimination. These two algorithms serve to factor products of two primes. Runtime between the two algorithms will be compared and analyzed, then runtime of quadratic sieve will be analyzed when the factors of the products have different bit lengths. \\
The contribution of this work has is summarized as follows:
\begin{itemize}
  \item Our experiment shows that pollard-rho performs better than quadratic sieve for numbers under 80 bits most of the time.
  \item We do an extensive analysis on the runtime of quadratic sieve under different circumstances- the bit difference of the two factors are large, and the bit difference of the two factors are small. Our result shows that the average runtime of the quadratic sieve tends to be shorter for products of the same bit length when the bit difference between the two factors is smaller. We verified that the quadratic sieve does better when the bit difference of the two factors are small.
\end{itemize}

\section{Related Work}
In this section, we briefly review the work related to pollard rho and quadratic sieve. In 1983, Joseph L. Gerver implemented quadratic sieve algorithm(QS), continued algorithm(CF) of Brillhart and Morrison and continued algorithm with early abort modification(CFEA) ~\cite{cfa}. In his work, he factored a 47-digit number that had never been factored into three primes using the quadratic sieve, and that number is 17674971819005665268668200903822757930076\\11. He also compared the runtime of QS, CF, and CFEA. It turns out that QS starts to do better than CF when the product exceeds 40 bits, and QS does better than CFEA when the product exceeds 60 bits. Although QS could beat CF and CFEA easily, when the product bit length exceeds 60 bits, pollard-rho is considered better than QS theoretically for products under 100 bits.

Peter Montgomery, an American mathematician who worked at the System Development Corporation and Microsoft Research, then did a modification to quadratic sieve and named it $\textbf{Multiple Polynomial Quadratic Sieve}$. Robert D. Silverman later implemented this modification of quadratic sieve and factored 45 digit numbers in 0.25 hours and 82 digit numbers in 1265 hours~\cite{multiple}. Before Silverman's time, there were only two implementations of the quadratic sieve algorithm. The second implementation, done from the 'Cunningham Project', used a Cray XMP supercomputer to factor the number ($10^{71}$ - 1)/9 (about 70 digits) in 9.5 hours ~\cite{multiple}. By the year 1994, the quadratic sieve was able to a 129- digit RSA number ~\cite{tale}. As better supercomputers and parallel machines are made, the quadratic sieve can factor numbers with more digits. 

$\textbf{Pollard-Rho factoring algorithm}$. Aminudin et al. analyzed the runtime of pollard-rho. Their experimental results show that their algorithm could factor a 44-bit number- 11752700814259- at 7.394 seconds and a 66-bit number- 49808531654765413631- at 28 seconds. They concluded that pollard-rho was significantly faster than Fermat's factorization ~\cite{pollard}. As a comparison, our pollard-rho algorithm can factor 11752700814259 in 0.0007 seconds and 49808531654765413631 in 0.06 seconds.

\section{Methodology}
We first show the theories behind pollard rho and our implementation of it. Then we show the mathematical steps required to implement the quadratic sieve and what techniques we incorporate into optimizing the quadratic sieve algorithm.

\subsection{Pollard Rho Factorization} The pollard-rho factoring algorithm is based on a probabilistic method to factor composite numbers N by finding the greatest common divisor(gcd) between the difference of two random numbers x and y (x and y are between 1 and N - 1) generated by an arbitrary function and N iteratively. The hope is that we could somehow find a pair of random numbers x and y such that the gcd(x - y, N) is the factor of N. The method was published by J.M. Pollard in 1975 and is based on the Birthday Paradox problem ~\cite{birthday}. The Birthday Paradox states that if we choose m samples from N items with replacement, with m to be large enough, we will choose some items twice. The pollard-rho algorithm uses the following sequences to generate random x and y values for finding the gcd of x-y and N:

\begin{center} $x_0$ $\leftarrow$ random positive integer  (mod $N$) \end{center}

\begin{center} c $\leftarrow$ random positive integer (mod $N$) \end{center}

\begin{center} $x_i$ $\leftarrow$ $f_c(x) = x_{i - 1} * x_{i - 1} + c$ (mod $N$)  \end{center}

\begin{center} $y_0$ $\leftarrow$ $f_c(x)$ (mod $N$)\end{center}

\begin{center} $y_i$ $\leftarrow$ $f_c(f_c(y_{i - 1}))$ (mod $N$)  \end{center}

Since the sequence can be at most $N-1$, the sequence will finally become periodic when the sequence gets larger. We find the nontrivial factor of the composite if $gcd(x-y, N)$ $\neq$ 1 and 
$gcd(x-y, N)$ $\neq$ $N$. According to the Birthday Paradox, the time to find a nontrivial factor is proportional to the size of $N$\footnote{\url{https://www.cs.umd.edu/users/gasarch/COURSES/456/F20/lecfactoring/bday.pdf}}. The expected number of steps to find the factor is approximate $N^{\frac{1}{4}}$. The runtime is considered extremely good with one flaw: the algorithm stops if it cannot find a nontrivial factor.

The implementation of our algorithm is shown below:

\begin{algorithm}
\caption{Pollard-Rho Algorithm}\label{alg:cap}
\begin{algorithmic}

\State $c \gets rand(1, N-1)$
\State $f_c(x) \gets x * x + c$ 
\State $x \gets rand(1, N-1)$ 
\State $y \gets f_c(x)$ (mod $N$)

\While{True}
    \State $x \gets f_c(x)$
    \State $y \gets f_c(f_c(y))$  

    \State $d \gets gcd(x-y, N)$
    \If{$d$ $\neq$ 1 and $d$ $\neq$ N}
    \State return $d$
    \EndIf
  
\EndWhile
\end{algorithmic}
\end{algorithm}

Besides the pseudocode shown above, we also added two extra tricks into the algorithm. We first check whether $N$ is prime or not, then we run the algorithm to prevent an infinite loop. Before running the main loop in the algorithm above, we check whether $N$ is divisible by the first ten primes to speed up the algorithm.

\subsection{Quadratic Sieve Algorithm}
The basic quadratic sieve algorithm is a more complicated factoring algorithm that contains several parts. 

$\textbf{B Smooth Numbers}$ Our goal is to factor a number $N$, which composes of two primes. One particular step of the quadratic sieve is to break down the numbers into smaller parts and see whether we can factor a large composite with smooth numbers. First, we want to choose a smooth bound $B$, a set of primes less than $B$, denoted as $\pi(B)$.

$\textbf{Parameter M}$ Next step is to use sieving to find a set of $a_i$ such that $b_i^2 \equiv a_i(mod N)$. In our implementaion of the quadratic sieve, we set $\sqrt{N}$ as our $b_1$, then we find $a_1$ such that $b_1^2 \equiv a_1(mod N)$. We repeat the process and update $b_{i+1}^2 = b_i^2 + 1$ until we reach $b_M^2 \equiv a_M(mod N)$.

$\textbf{Forming the Matrix}$ After we find the set of $a_i$ from the previous step, we want to write each $a_i$ as prime factors from the B smooth numbers we found in step one and generate exponent vectors for factors of each $a_i$ then mod them by 2. Example:

The B smooth bound we choose is 15, so the set of primes less than 10 is {2, 3, 5, 7}. The number we want to factor is 400289. Then $\sqrt{400289}$ = 633. We compute 633$\equiv 2^4\cdot5^2$(mod 400289). The vector we generate for $2^4\cdot5^2$ is $\vec{a_1}$ = (4, 0, 2, 0) (mod 2) = (0, 0, 0, 0). 

We form a 2-dimensional matrix from a set of exponent vectors generated from all the $a_i$, then use the Gaussian elimination process to find a linear combination of rows mod 2 that could sum up to the 0 vector.

$\textbf{Finding the Factors}$ Once we found an identity such that $b^2\equiv a^2(mod N)$, we can rewrite the identity to be $(b-a)(b+a)\equiv 0 (mod N)$. The final step is to find the greatest common divisor between $(b-a)$ and $(b+a)$. If the greatest common divisor between $(b-a)$ and $(b+a)$ is not the trivial factor- 1, we find a factor for $N$. If it is the trivial factor, we increase our $B$ smooth bound and M parameters and repeat step one. Algorithm 2 shows the basic algorithm for the quadratic sieve.

\begin{algorithm}
\caption{Quadratic Sieve Algorithm}\label{alg:cap}
\begin{algorithmic}

\State Given $N, B, M$
\State Generate a list of primes less than $B$
\State $b_1 \gets \sqrt{N}$ 
\State An empty matrix list $A$
\For{$k \gets 1$ to $M$}                    
\State $a_k = b_k $ (mod $N$)
\State $c_i$ $\gets$ exponent vectors of $a_i$ (mod 2)
\State Add $c_i$ to $A$
\EndFor

\State Perform Gaussian elimination on A
\State Let a, b be a list of linear combination of rows in A such that  $a^2 \equiv b^2 (mod N)$

\If{$gcd(a-b, a+b)$ $\neq$ 1 }
    \State return $gcd(a-b, a+b)$
\Else
    \State $B$ $\gets$ $B$ + 10
    \State $M$ $\gets$ $M$ + 100
    \State Repeat the process again
\EndIf

\end{algorithmic}
\end{algorithm}

The asymptotic running time for quadratic sieve is  $O(e^{\sqrt{1.125ln(N)ln(ln(N))}}))$ ~\cite{quad_runtime}.

\subsection{Safe Prime Generators}
\hspace{2.5 mm}We want to generate random primes of certain lengths, but sometimes it is expensive to check whether a large number is prime or even impossible. Thus, we want to generate numbers considered primes with high probabilities but can still be composites with a low probability.

\section{Experiments}
In this section, we introduce the experimental setup and present the results of our experiment. Concrete and specific examples will be presented to give a better understanding of our results. 

\subsection{Experimental Setup}
\textbf{Dataset} We want to test the performance of the two algorithms on composites with different bit lengths. We used the safe prime method to generate safe primes of different bit lengths- 40 bits, 50 bits, and 60 bits. We also want to compare the performance of both algorithms for composites of certain bit lengths but with prime factors of different bit lengths. An example for generating 40-bit composites is a combination of 5 bit and 35-bit primes multiplied together to get a 40-bit product. The list of combinations of primes we generated is shown in table 1. In addition, we want to examine the two algorithms' performance on composites of random bit lengths (The bit lengths of the composites are not necessarily multiples of 5). Thus, we generated 4462 pairs of primes with random bit lengths, but the products of those pairs of primes do not exceed 70 bits. The experiment is run in a personal laptop MacBook pro with a RAM of 16 GB and 2GHz Quad-Core Intel Core i5. No parallel machine nor GPU is used in the whole experiment.

\begin{table}
\centering
\begin{tabular}{c|c|c}
 \hline
Composite Size & Prime 1 Size & Prime 2 Size\\ [0.1ex] 
 \hline\hline
40 &  5 & 35 \\
40 & 10 & 30 \\
40 & 15  & 25 \\
40 & 20  & 20 \\
50 &  5 & 45 \\
50 &  10 & 40 \\
50 &  15 & 35 \\
50 &  20 & 30 \\
50 &  25 & 25 \\
60 &  5 & 55 \\
60 &  10 & 50 \\
60 &  15 & 45 \\
60 &  20 & 40 \\
60 &  25 & 35 \\
60 &  30 & 30 \\
 \hline
\end{tabular}
\caption{\textnormal{Composites generated by combination of prime factors of different bit lengths}}
\end{table}

\subsection {Bit Length of Products and Runtime of the Two Algorithms Comparison on Random Bit Composites}
First, we want to compare the performance of quadratic sieve and pollard-rho on composites of random bit lengths below 70 bits. In this experiment, we limit the runtime of finding suitable B and M parameters for the quadratic sieve algorithm to 3 minutes for the sake of limited time for this project. We compare the runtime of the two algorithms on products in different bit lengths. We generated 4462 products of random bit lengths and tested the runtime of both programs. Figure 1 visualizes the runtime of the quadratic sieve algorithm on products of random bit lengths. Figure 2 shows the runtime of pollard rho algorithms for the same set of products. Among 4462 products, only 4268 products are being successfully factored in the time limit. From the two graphs, we see that the maximum runtime of the quadratic sieve is about 1.2 seconds.

On the contrary, the highest runtime for pollard rho on the products is about 0.175 seconds. Thus, we see that pollard rho completely beats quadratic sieve on random bit length products no greater than 70 bits. In addition, it seems to take the quadratic sieve the longest time to factor products of around 65 bit length. The runtime of the quadratic sieve seems to increase as the bit length of the products increase, which looks reasonable. However, some numbers between bit length of 50 bits and 60 bits took significantly longer runtime than other numbers smaller than 60 bits. Here, we consider the time to factor a number smaller than 60 bits more than 0.6 seconds to be long. Table 2 shows the numbers smaller than 60 bits but took more than 0.6 seconds to factor. Among those products, the performance of pollard-rho is still very good- all below 0.001 seconds. 

\begin{table}
\centering
\begin{tabular}{c|c|c|c|c|c}
 \hline
Product & Factor 1 & Factor 2 & Factor 1 Length & Factor 2 Length & Pollard Runtime\\ [0.5ex] 
 \hline\hline
27522003582288223	& 2963	& 9288560102021	& 12 &	44 & 0.0000472\\
1751357076372217 &	4441 &	394360971937 &	13 &	39 & 0.0001149\\
349739602979093	& 291148573	& 1201241	& 29 &	21 & 0.0007229\\
185456974731188183	& 18796061 &	9866800003 &	25 & 34 & 0.0002370\\
 \hline
\end{tabular}
\caption{\textnormal{Products that take quadratic sieve more than 0.6 seconds to run}}
\end{table}

For Pollard rho, it is not surprising that the program's runtime gradually increases as the bit length of the products increases. As the number gets larger, the number of cycles to hit the correct candidate factors becomes bigger. There is a significant jump in the runtime of Pollard rho from 60 bit to 70 bits.

\begin{figure}
\centering
\includegraphics[width=0.8\linewidth]{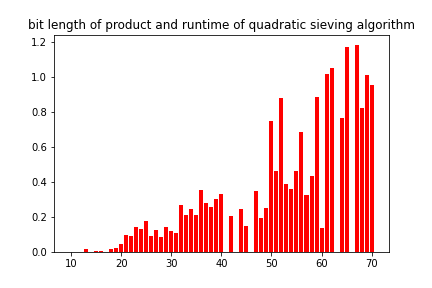}

\caption{The histogram shows the bit length of composites and the runtime corresponding to that bit length for quadratic sieving algorithm}
\end{figure}

\begin{figure}
\centering
\includegraphics[width=0.8\linewidth]{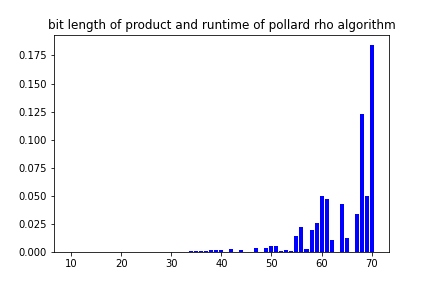}

\caption{Bit length of composites and the runtime corresponding to that bit length for Pollard $\rho$}
\end{figure}

\subsection{Analysis on Products Quadratic Sieve Performs Better than Pollard-Rho} 

The following section compares the runtime between quadratic sieve and pollard rho for products of 40 bits, 50 bits, and 60 bits. 

We randomly generated 1276 products of 40 bits, 2749  products of 50 bits, and 3561 products of 60 bits. We then factored them using both algorithms. The result shows that the runtime of pollard-rho is faster than the quadratic sieve for all 40-bit products. Two numbers quadratic sieve performs better than pollard-rho for 50-bit products, and five numbers quadratic sieve is better than pollard-rho for 60-bit products. Table 3 lists the products that quadratic sieve does better than pollard-rho. From the table, we see that the quadratic sieve does better than pollard-rho for all the products. The bit length of factor 1 and factor 2 are the same. The average runtime for quadratic sieve on products of two 25-bit primes is 0.140562 seconds and 0.004899 seconds for pollard-rho. The average runtime for quadratic sieve on products of two 30-bit primes is 0.224792 seconds and 0.047880 seconds for pollard-rho. 
For all the products in table 3 except the one with factors 633488353 and 674203633, the runtime of pollard-rho is above the average runtime of pollard rho on products with factors of the same bit length. On the other hand, the runtime of the quadratic sieve on those products is much smaller than the average runtime on other 25-25 bit products and 30-30 bit products. Thus, we conclude that the quadratic sieve performs better than pollard-rho on products of the same bit length when pollard-rho performs below average, and the quadratic sieve performs above average.

\begin{table}
\centering
\begin{tabular}{c|c|c|c|c|c}
 \hline
Factor 1 & Factor 2 & Factor 1 Length & Factor 2 Length & Pollard Runtime & Quad Runtime\\ [0.5ex] 
 \hline\hline
17410997 & 19124933 & 25 &	25 & 0.0147488 &0.0138339996  \\
21094421 &	21577789 &	25 & 25 & 0.0104671 & 0.0087039470\\
642478871	& 674265899	& 30 &	30 & 0.0537941 & 0.0379431247\\
630635641 &	563503649 &	30 & 30 & 0.058249 & 0.0529510974\\
718778023 &	648638647 &	30 & 30 & 0.0558028 & 0.0537159442\\
633488353 &	674203633 &	30 & 30 & 0.0393589 & 0.0285172462\\
639632243 &	758675243 &	30 & 30 & 0.053247 & 0.0451288223\\
 \hline
\end{tabular}
\caption{\textnormal{Products quadratic does better than pollard-rho}}
\end{table}

\subsection{Failure Rate of Quadratic Sieve for 40-bit, 50-bit, and 60-bit Products Within Three Minutes}
Due to the time limits of this experiment, it is impossible to allow the program to run an infinite amount of time to factor some numbers that are hard to factor. Thus, if it takes the program more than three minutes to find the suitable B and M values and factor a product, we will cut off the program and factor the next product.

$\textbf{Failures on 40-bit Products}$
There are 14 numbers quadratic sieve cannot factor for 40-bit products. The numbers the program cannot factor is shown in table 3 in the appendix. From table 4, we see that all the products quadratic sieve cannot factor are composed of a 10-bit prime factor and a 30-bit prime factor.

\begin{table}
\centering
\begin{tabular}{c|c|c}
 \hline
Product & Prime 1 Size & Prime 2 Size\\ [0.1ex] 
\hline\hline
369062373143 &  10 & 30 \\
534106988197 & 10 & 30 \\
508085497589 & 10  & 30 \\
369901521617 & 10  & 30 \\
369901521617 &  10 & 30 \\
458343342553 &  10 & 30 \\
341238699311 &  10 & 30 \\
421135689817 &  10 & 30 \\
563755932497 &  10 & 30 \\
482560209653 &  10 & 30 \\
477265583519 &  10 & 30 \\
456886122281 &  10 & 30 \\
442321850393 &  10 & 30 \\
386706654493 &  10 & 30 \\
 \hline
\end{tabular}
\caption{\textnormal{40-bit products QS cannot factor within three minutes}}
\end{table}

$\textbf{Failures on 50-bit Products}$
For 50-bit products, there are 206 products quadratic sieve cannot factor within three minutes. Table 5 shows the bit length of two primes and the number of products that cannot be factored on time. From table 4, we see that the most number of products that cannot be factored in are combinations of 10-bit primes and 40-bit primes. We see products that contain 10-bit primes are harder to factor.

\begin{table}
\centering
\begin{tabular}{c|c|c}
 \hline

Prime 1 Size & Prime 2 Size & Failure Count\\ [0.1ex] 
\hline\hline
5 &  45 & 38 \\
10 & 40 & 51 \\
15 & 35  & 38 \\
20 & 30  & 37 \\
25 &  25 & 42 \\
 \hline
\end{tabular}
\caption{\textnormal{Counts on products that cannot be factored for different combination of prime factors on 50-bit products}}
\end{table}

$\textbf{Failures on 60-bit Products}$
Table 6 shows the number of 60-bit products that cannot be factored. However, for 60-bit products, the most of products that cannot be factored is not a combination of a 10-bit prime and a 50-bit prime. It is a combination of a 5-bit prime and a 55-bit prime. A possible reason why products cannot be factored is the combination of bit lengths between the two prime factors. A hypothesis is that when the difference of bit lengths of two primes is large, it takes more time for the quadratic sieve to factor those numbers. We will verify this hypothesis in the following section.

\begin{table}
\centering
\begin{tabular}{c|c|c}
 \hline
Prime 1 Size & Prime 2 Size & Failure Count\\ [0.1ex] 
\hline\hline
5 &  55 & 417 \\
10 & 50 & 383 \\
15 & 45  & 367 \\
20 & 40  & 383 \\
25 &  35 & 373 \\
30 &  30 & 406 \\
 \hline
\end{tabular}
\caption{\textnormal{Counts on products that cannot be factored for different combination of prime factors on 60-bit products}}
\end{table}

\begin{table}
\centering
\begin{tabular}{c|c|c|c}
 \hline
Bit Difference & Factor 1 Bit Length & Factor 2 Bit Length & Percentage that Can be Factored\\ [0.1ex] 
\hline\hline
30 &  5 & 35 & 1.0000 \\
20 & 10 & 30 & 0.9646 \\
10 & 15  & 25 & 1.0 \\
0 & 20  & 20 & 1.0 \\
40 &  5 & 45 & 0.9023\\
30 &  10 & 40 & 0.9136 \\
20 & 15 & 35 & 0.9354 \\
10 & 20 & 30 & 0.9370 \\
0 & 25 & 25 & 0.9294 \\
50 & 5 & 55 & 0.2992 \\
40 & 10 & 50 & 0.3541 \\
30 & 15 & 45 & 0.3822 \\
20 & 20 & 40 & 0.3541 \\
10 & 25 & 35 & 0.3721 \\
0 & 30 & 30 & 0.3142 \\
 \hline
\end{tabular}
\caption{\textnormal{Shows the percentage of successful factoring within three minutes for products of different big lengths and different bit length between the two factors of the product}}
\end{table}

\subsection{Percentage of Products that Cannot Be Factored For 40-bit, 50-bit and 60-bit Products}
Because the number of samples for 40-bit, 50-bit and 60-bit products are different, looking at how many products quadratic sieve cannot factor for each bit length is insufficient. We must look at the percentage of products quadratic sieve cannot factor given the bit length of the product. The result shows that for 40-bit products, about 1.097 percent of the numbers cannot be successfully factored within three minutes. For 50-bit products, about 9.657 percent of the numbers cannot be factored within three minutes. For 60-bit products, the failure rate is up to 65.403 percent. Thus, the longer the bit length of the product is, the harder it is for the quadratic sieve to find out the suitable B and M values and factor the products on time. When the size of a product gets larger, the algorithm may spend most of its time finding the correct B and M values instead of doing the actual process of factoring the number, such as Gaussian elimination.

\subsection{How Well Does Quadratic Sieve Perform When the Bit Difference of Primes Changes}
Table 7 shows how the quadratic sieve performs when the bit difference of the two prime factors vary. Especially, they show the percentage of numbers that can be factored within three minutes for different combinations of bit length primes.

From table 7, we see that except for 40-bit products, the success rate of factoring in three minutes seem to be the lowest when the bit difference of the two primes are the largest. As the bit difference gets smaller, the success rate of factoring increases for 50-bit and 60-bit products.

\subsection{Comparing Average Runtime for Quadratic Sieve when the Difference of Bit Lengths for the Two Primes are large and small}
The previous section concluded that the percentage of successful factoring within three minutes for the quadratic sieve to factor a number at a certain bit length is the lowest when the bit difference of the two primes are large. Thus, we examine whether changing the bit difference between two factors for fixed bit length products can affect the average runtime of the quadratic sieve. Table 8 shows the average runtime for the quadratic sieve for combinations of products of different bit lengths and different bit difference

Indeed, from table 8, we see that for 40-bit, 50-bit and 60-bit products, the average runtime in seconds to factor a product is the highest when the bit difference is the greatest. Thus, the runtime of the quadratic sieve depends not only on the bit length of the product but also on the bit difference between two prime factors.~\cite{gaussian}.

\begin{table}
\centering
\begin{tabular}{c|c|c}
 \hline
Product Size in Bits & Bit Difference & Average Runtime in Seconds \\ [0.1ex] 
\hline\hline
40 &  30 & 0.123 \\
40 & 20 & 0.122 \\
40 & 10  & 0.096 \\
40 & 0  & 0.098 \\
50 &  40 & 0.245 \\
50 &  30 & 0.154 \\
50 &  20 & 0.146 \\
50 & 10 & 0.143 \\
50 & 0 & 0.140 \\
60 & 50 & 0.242 \\
60 & 40 & 0.233 \\
60 & 30 & 0.240 \\
60 & 20 & 0.236 \\
60 & 10 & 0.242 \\
60 & 0 & 0.224 \\
 \hline
\end{tabular}
\caption{\textnormal{Shows the average runtime in seconds for different size products}}
\end{table}

\section{Limitations}

\textbf{Computer Resources}
We have limited computing power for running the two algorithms. The only computer we use is a mac book pro with 16GB memory. Quadratic sieve algorithm would speed up if it runs in a parallel machine since it can perform Gaussian elimination simultaneously on multiple rows in a matrix. However, for a normal personal computer, the quadratic sieve algorithm takes a long time to find the correct B and M values. With the time limit, the success rate of factoring a number for a quadratic sieve becomes much lower when the bit length of the product exceeds 70 bits. Thus, we did not run quadratic sieve and pollard rho on numbers greater than 70 bits. However, in theory, the quadratic sieve should do better than pollard rho when the product is over 100 bits. We ran an experiment on both algorithms for an 80-bit number, the runtime of pollard rho is about 22.95 seconds, and the runtime of the quadratic sieve is about 44.25 seconds. Afterwards, we ran both algorithms on a 100-bit number, and pollard rho did not successfully factor in ten minutes. Quadratic sieve factored the number in 587.66 seconds. Although there is only one example for 100-bit numbers, we at least find an example that satisfies the theory.

\textbf{Time Limits} Although the time limit of three minutes for quadratic sieve and pollard rho greatly saves our time on factoring thousands of examples, we did not record the time for finding the correct B and M values for quadratic sieve, which is useful. In many cases, the quadratic sieve cannot factor the products merely because it cannot find the suitable B and M values on time. If we have a better algorithm for finding the correct B and M values for the quadratic sieve, the success rate of factoring numbers of quadratic sieve would be much higher.

\textbf{Limited Sample Sizes} Because we can only analyze 40-bit, 50-bit and 60-bit products, the comparisons we can do for pollard rho and quadratic sieve are very limited. We cannot see a significant performance of quadratic sieve over pollard rho for 100-bit length numbers for each pair of B and M values. It would take 10 minutes for the quadratic sieve to factor and maybe more than 10 minutes for the pollard rho to factor. The time cost of doing large bit length numbers is very high. Also, the failure rate of factoring 80-bit numbers for quadratic sieve is high for a three-minute timing- about 90 percent failure rate. The limited sample size limits the type of comparison we want to do between the two algorithms.

\section{Conclusion}
In this paper, we analyzed the performance of quadratic sieve and pollard-rho from various aspects. Pollard-rho dominates the performance for composites smaller than 80 bits. However, we are still able to find some composites that the quadratic sieve outperforms pollard-rho. We also conclude that quadratic sieve is more likely to succeed when the bit difference between the factors is small but with the same bit size products. Also, the average performance of the quadratic sieve increases for composites of the same bit size but smaller bit difference. 

\section{Summary and Outlook}
\subsection{Time Span to run the algorithms}Increase the timespan for factoring in each product. It takes 587.6639738 seconds- almost 10 minutes for the quadratic sieve to factor the product. Also, we have to count for the time to find out the B and M values, which means 20 minutes to factor each product. The time to run the algorithms might take a month.

\subsection{Factor more specific bit length primes}Find how much time it takes to factor a 40-bit number when the two products are 5 bit and 35 bit, 6 bit and 34 bit, 7 bit and 33 bit, etc. Do the same for 45 bit, 50 bit, 55-bit, 60-bit products.

\subsection{Finding a better algorithm for finding B and M values} Instead of incrementing B and M by ten each alternately, starting B = 10 and M = 10, for larger numbers, we start B and M with 100 and 1000 and increment the two values by 100 or other intervals alternately.

\bibliographystyle{unsrtnat}
\bibliography{references}

\end{document}